\documentclass[aps,prl,twocolumn,tightenlines,superscriptaddress]{revtex4}
\usepackage{epsfig}
\usepackage{graphicx}
\usepackage{subfig}
\usepackage{psfrag}
\usepackage{amsmath,amssymb}
\usepackage{colordvi}
\usepackage{color}
\usepackage{amsfonts}
\usepackage{enumerate}
\usepackage{slashed,bm}

\begin{document}

\title{Spin Decomposition of Electron in QED}

\author{Xiangdong Ji}
\affiliation{INPAC, Department of Physics and Astronomy, Shanghai Jiao Tong University, Shanghai, 200240, P. R. China}
\affiliation{Maryland Center for Fundamental Physics, University of Maryland, College Park, Maryland 20742, USA}
\author{Andreas Sch\"afer}
\affiliation{Institut f\"ur Theoretische Physik, Universit\"at Regensburg, \\
D-93040 Regensburg, Germany}
\author{Feng Yuan}
\affiliation{Nuclear Science Division, Lawrence Berkeley National Laboratory, Berkeley, CA, 94720}
\author{Jian-Hui Zhang}
\affiliation{Institut f\"ur Theoretische Physik, Universit\"at Regensburg, \\
D-93040 Regensburg, Germany}
\author{Yong Zhao}
\affiliation{Maryland Center for Fundamental Physics, University of Maryland, College Park, Maryland 20742, USA}
\affiliation{Nuclear Science Division, Lawrence Berkeley\\
National Laboratory, Berkeley, CA, 94720}

\date{\today}
\vspace{0.5in}
\begin{abstract}
We perform a systematic study on the spin decomposition of an electron in QED at 
one-loop order. It is found that the electron orbital angular momentum defined in 
Jaffe-Manohar and Ji spin sum rules agrees with each other, and the so-called 
potential angular momentum vanishes at this order.
The calculations are performed in both dimensional regularization and Pauli-Villars 
regularization for the ultraviolet divergences, and they lead to consistent results.
We further investigate the calculations in terms of light-front wave functions, 
and find a missing contribution from the instantaneous interaction in light-front quantization. 
This clarifies the confusing issues raised recently in the literature on the spin 
decomposition of an electron, and will help to consolidate the spin physics program for nucleons
in QCD.
\end{abstract}

\maketitle

\section{Introduction}

One of the most important goals in hadron physics is to understand the longitudinal spin structure of the proton. Although the total angular momentum of an isolated system is well-defined in QCD, apportioning it into the spin and orbital angular momentum (OAM) contributions of the quarks and gluons has been a challenging task. Over the past twenty-five years, two well-known sum rules have been proposed and inspired much theoretical and experimental effort to understand the longitudinal spin structure of the proton. 

One of them is the sum rule proposed by Jaffe and Manohar~\cite{Jaffe:1989jz},
\begin{equation}
{1\over 2} = {1\over2}\Delta\Sigma  + \mathcal{L}_q + \Delta G + \mathcal{L}_g\ ,
\end{equation}
where each individual term is defined to be the proton matrix element of the canonical spin and OAM tensors in the infinite momentum frame (IMF) or on the light-front:
\begin{eqnarray}
M^{+12} &= & {1\over2}\bar{\psi}\gamma^+\gamma^5 \psi + \bar{\psi}\gamma^+(\vec{x} \times (-i\vec{\nabla}))^3\psi \nonumber\\
&&+ \epsilon^{+-ij} \mbox{Tr} F^{i+}A^j + 2\mbox{Tr} F^{i+} (\vec{x} \times \vec{\nabla})^3 A^i \ .
\label{jm}
\end{eqnarray}
Here $i, j=1,2$, and a summation over dummy indices is implied. Except for the quark spin, all the other terms are gauge variant and have to be fixed in the light-cone gauge $A^+=(A^0+A^3)/\sqrt{2}=0$. The residual gauge invariance is fixed by imposing anti-periodic boundary conditions $A^{1,2}(\vec{x}_\perp, \infty^-) = - A^{1,2}(\vec{x}_\perp, \infty^+)$. Despite its gauge dependence, there is a strong motivation behind the Jaffe-Manohar sum rule since both $\Delta \Sigma$ and $\Delta G$ are measurable in high-energy scattering experiments~\cite{deFlorian:2009vb,deFlorian:2014yva}.

The other sum rule proposed by Ji~\cite{Ji:1996ek} is frame independent and manifestly gauge invariant,
\begin{equation}
{1\over 2} \ = \ J_q + J_g \ = \  {1\over2}\Delta\Sigma + L_q + J_g\ ,
\end{equation}
where the quark spin, the kinetic quark OAM, and the total gluon angular momentum are the proton matrix elements of each term in the following expression:
\begin{equation}
M^{012} = {1\over2}\bar{\psi}\gamma^+\gamma^5 \psi + \bar{\psi}\gamma^+(\vec{x} \times (-i\vec{D}))^3\psi + [\vec{x}\times (\vec{E}\times\vec{B})]^3\ ,
\end{equation}
with $\vec{D} = \vec{\nabla} - ig\vec{A}$. The quark and gluon angular momenta also satisfy separate sum rules,
\begin{equation}
J_{q,\,g} = {1\over 2} \left[ A_{q,\,g}(0) + B_{q,\,g}(0)\right] \ ,
\end{equation}
where $A_{q,\,g}$ and $B_{q,\,g}$ are the gravitational form factors of the symmetric quark and gluon energy-momentum tensors. 
Therefore, the quark OAM in the Ji sum rule can be obtained by
\begin{equation}
L_q=\frac{1}{2}\left[A_q(0)+B_q(0)\right]-\frac{1}{2}\Delta\Sigma \ .\label{ji-oam}
\end{equation}
The Ji sum rule has received considerable attention for its relation to generalized parton distributions (GPDs) and experimental probes~\cite{Ji:1996ek, Ji:1996nm, Hoodbhoy:1998yb}. 


There have been investigations on the difference between the canonical and kinetic OAM in the Jaffe-Manohar and Ji sum rules. According to their definition, $\mathcal{L}_q$ and $L_q$ are simply related by~\cite{Ji:2012ba,Hatta:2012cs}
\begin{equation}
L_q = \mathcal{L}_q + l_{\text{pot}} \ ,
\label{relation}
\end{equation}
where $l_{\text{pot}}$ is the proton matrix element of the potential angular momentum 
\begin{equation}
l^z_\text{pot} = - e \bar{\psi} \gamma^+ (\vec{x} \times \vec{A})^3 \psi\ .
\end{equation}
While attempts have been made to compute $L_q$ using lattice QCD~\cite{Mathur:1999uf}, strategies for a direct lattice computation of $\mathcal L_q$ have been proposed only recently~\cite{Ji:2014lra,Zhao:2015kca}. Despite that a lattice evaluation of the latter is required to show the numerical significance of their difference, model-dependent analyses and the study of the situation in QED can already shed some light on this problem. In fact, few people will trust any QCD analysis unless the QED case is completely understood. 
In Ref.~\cite{Burkardt:2008ua}, Burkardt et al. computed the quark OAM in both the Jaffe-Manohar and Ji sum rules in a scalar diquark model and found an agreement. This is expected since the scalar diquark model does not contain a gauge field. They also computed the electron OAM to $\mathcal O(\alpha)$ in QED using the formalism of light-front wave functions, and found a finite difference between the electron OAM defined by Jaffe-Manohar and Ji decompositions. This also implies a non-vanishing finite difference in the quark case in QCD~\cite{Burkardt:2008ua}. Burkardt~\cite{Burkardt:2012sd} interpreted this difference as the change in the quark OAM as it leaves the target in a deep-inelastic scattering (DIS) experiment, which is similar to the mechanism responsible for the transverse single-spin asymmetries in semi-inclusive DIS. 

In a recent paper~\cite{Liu:2014fxa}, Liu and Ma also computed the canonical and kinetic OAM for a dressed electron to $\mathcal O(\alpha)$ in QED. Their result confirmed the finite difference obtained by Burkardt et al.~\cite{Burkardt:2008ua}, if the same ultraviolet (UV) regularization (Pauli-Villars) is adopted. Moreover, they considered the contributions from the potential angular momentum and surface terms arising from the derivation of the Ji sum rule, and found that both have a non-vanishing anomalous dimension at $\mathcal O(\alpha)$. They then claimed that the difference between the canonical and kinetic OAM cannot be explained by the potential angular momentum or the surface term, and that the Ji sum rule is incorrect for quarks and gluons separately.


In this paper, we show that to $\mathcal O(\alpha)$ in QED, the electron OAM defined in the Jaffe-Manohar and Ji sum rules are consistent, and actually equal to each other. The fact that the result of Liu and Ma for the potential angular momentum is incorrect can be seen from the following: as was shown in Refs.~\cite{Ji:1996ek,Ji:1995cu}, the potential angular momentum does not contribute to the evolution at one-loop order, therefore it must have a vanishing anomalous dimension (this conclusion applies to both QCD and QED, since at one-loop order the contributing diagrams in QCD are Abelian-like). Actually, our explicit calculation shows that the contribution of the potential angular momentum vanishes at $\mathcal O(\alpha)$, so do the surface terms. Moreover, as we will point out, in light-front quantization a proper implementation of UV regularization is of crucial importance to resolve the inconsistency between the canonical and kinetic electron OAM in Refs.~\cite{Burkardt:2008ua, Liu:2014fxa}. This is closely related to the subtle difference between light-front and equal-time quantization, as the Hamiltonian of the former contains an instantaneous interaction while the latter does not. If this difference is properly taken into account, the electron OAM in the Jaffe-Manohar and Ji sum rules are equal at $\mathcal O(\alpha)$ (Note that this does not mean that there is no difference between them to all orders, because the potential angular momentum might contribute starting from $\mathcal O(\alpha^2)$.). We also compute the same quantities using the Feynman diagram approach, where the electron OAM in the Jaffe-Manohar and Ji sum rules are found to be equal to each other at $\mathcal O(\alpha)$.

The rest of the paper is organized as follows. In Sec.~II, we compute the electron OAM at one-loop order, with two regularization methods---dimensional regularization and Pauli-Villars regularization---for the UV divergences in the calculations. In Sec.~\ref{oamlfwf}, we review the computation of the electron OAM in the Jaffe-Manohar and Ji sum rules in Ref.~\cite{Burkardt:2008ua}, and explain why the difference between them should be zero. 
We then conclude in Sec.~\ref{concl}. Some of the detailed derivations are given in the Appendix.


\section{Electron Orbital Angular Momentum at One-loop Order}

We follow the procedure of Refs.~\cite{Ji:1996ek,Ji:1995cu} to calculate the electron spin and OAM at one-loop order in QED. Different from Refs.~\cite{Ji:1996ek,Ji:1995cu}, in order to obtain the finite contributions
at this order, we will keep a non-zero electron mass $m$ in the calculations, which also 
regularizes the collinear divergences in the calculations.

\subsection{Dimensional regularization}

At leading order, the total spin is carried by the electron spin
$\Delta\Sigma_e/2$. This will be modified by photon radiation at one-loop order, 
where we have to deal with the UV divergences. In the following, we
will first apply dimensional regularization with $D=4-2\epsilon$, where we have to specify the prescription for 
$\gamma_5$. 
To perform the calculations consistently, we follow the HVBM scheme~\cite{'tHooft:1972fi}:
\begin{equation}
\gamma_5 \equiv {i\over 4!} \epsilon^{\mu\nu\rho\sigma}\gamma_\mu \gamma_\nu \gamma_\rho \gamma_\sigma \ .
\end{equation}
Splitting the $D$-dimensional metric tensor into its four- and $(D-4)$-dimensional components, we have
\begin{equation}
g^{\mu\nu} = \bar{g}^{\mu\nu} + \hat{g}^{\mu\nu} \ , \ \ \ \bar{g}^\mu_{~\mu} = 4, \ \ \ \ \hat{g}^\mu_{~\mu} =D-4=-2\epsilon\ ,
\end{equation}
and
\begin{eqnarray}
\{ \gamma^\mu, \gamma^5 \} &=& 0, \ \ \ \ \mbox{for }\mu=0,1,2,3,\nonumber\\
\left[ \gamma^\mu, \gamma^5 \right] &=& 0, \ \ \ \ \mbox{otherwise\ .}
\end{eqnarray}
It is straightforward to calculate electron spin at one-loop order, and we obtain
\begin{eqnarray}
\frac{\Delta\Sigma_e^{(1)} }{2} &=& \frac{e^2}{4\pi}\int dx\int \frac{d^{D-2}k_\perp}{(2\pi)^{D-2}}\left[\frac{1+x^2}{1-x}\frac{1}{\vec{k}_\perp^2+(1-x)^2m^2}\right.\nonumber\\
&&\left.-\frac{2(1-x)(1-x+x^2)m^2}{(\vec{k}_\perp^2+(1-x)^2m^2)^2}+\frac{3\epsilon(1-x)}{\vec{k}_{\perp}^2+(1-x)^2m^2}\right]\nonumber\\
&&+\Gamma^v\ ,\label{ez0}
\end{eqnarray}
where $\Gamma^v$ is the virtual contribution. To calculate $\Gamma^v$, we need the electron wave function renormalization constant in the light-cone gauge, a proper definition of which is given by (for a similar treatment in the axial gauge $A^z=0$, see e.g. Ref.~\cite{Xiong:2013bka}),
\begin{equation}
Z =1 + {1\over 2P^+} \bar{u}(P){\partial \Sigma(P)\over \partial P^-}  u(P) \ ,
\end{equation}
where $\Sigma(P)$ is the self energy of an on-shell electron. 
This definition guarantees that the vector current $\bar{\psi}\gamma^+\psi$ is 
strictly conserved (at least at one-loop order) and is consistent with the renormalization constant defined from light-front wave functions below.

The UV divergences from the
real and virtual contributions will be cancelled out between each other, and we
are left with a finite contribution. 
In the above equation, we have kept the mass-dependent term as well as
the $\epsilon$-term, because they will lead to a finite contribution in the end. 
Clearly, there is the splitting kernel in the above equation, and the $\epsilon$-term
is consistent with previous calculations (see, for example, Ref.~\cite{Vogelsang:1996im}). 
After applying dimensional regularization for the $k_\perp$ integral, 
we obtain the following result,
\begin{eqnarray}
\frac{\Delta\Sigma_e^{(1)} }{2}&=& \frac{\alpha}{4\pi}\int dx\left[\frac{1+x^2}{1-x}\left(N_\epsilon+\ln\frac{\mu^2}{(1-x)^2m^2}\right)\right.\nonumber\\
&&\left.-\frac{2(1-x+x^2)}{1-x}+3(1-x)\right]+\Gamma^v\ ,
\end{eqnarray}
where $N_\epsilon=1/\epsilon-\gamma_E+\ln(4\pi)$, and $\mu$ is the renormalization scale. As mentioned above,
the UV divergences are cancelled out between the real and virtual contributions,
and the final result is
\begin{eqnarray}
\frac{\Delta\Sigma_e^{(1)} }{2}&=& \frac{\alpha}{4\pi}\int dx\ 2(1-x)=\frac{\alpha}{2\pi}\cdot{1\over2}\ .\label{ez}
\end{eqnarray}
To calculate the electron OAM in the Ji sum rule, we need to subtract
the electron spin of Eq.~(\ref{ez}) from the sum rule of Eq.~(\ref{ji-oam}), for which we 
will calculate $A_e(0)$ and $B_e(0)$. The one-loop result for $B_e(0)$ has been calculated
in the literature, which is finite and does not depend on the regularization method. Here, we quote the result in~\cite{Burkardt:2008ua},
\begin{equation}
B_e(0)=\int d x\ xE(x)=\frac{\alpha}{2\pi}\int dx\ 2x^2=\frac{\alpha}{2\pi}\cdot\frac{2}{3} \ ,
\end{equation}
where $E(x)=E(x,0,0)$ is a GPD defined in Refs.~\cite{Ji:1996ek, Ji:1996nm}.
The one loop result for $A_e(0)$ is
\begin{eqnarray}
A_e^{(1)}(0)&=&\frac{e^2}{2\pi}\int dx\ x\int\frac{d^{D-2}k_\perp}{(2\pi)^{D-2}}\left[\frac{1+x^2}{1-x}\frac{1}{\vec{k}_\perp^2+(1-x)^2m^2}\right.\nonumber\\
&&\left.-\frac{2x(1-x)m^2}{(\vec{k}_\perp^2+(1-x)^2m^2)^2}-\frac{\epsilon(1-x)}{\vec{k}_{\perp}^2+(1-x)^2m^2}\right]\nonumber\\
&&+\Gamma^v\ ,
\end{eqnarray}
where the virtual contribution is the same as that for the electron spin in 
Eq.~(\ref{ez0}). Without the additional factor $x$ in the integral of $dx$,
the above equation gives the splitting of electron to electron at one-loop order in QED.
That provides an important cross check, because the integral of the electron
splitting function (over $x$) vanishes at any order of perturbative theory. 
Comparing the above equation to Eq.~(\ref{ez0}), there are two differences:
one is the mass term, and the other is the $\epsilon$-term. The difference
in the $\epsilon$-term is consistent with the corresponding splitting
functions for the unpolarized and polarized quark distributions
in Ref.~\cite{Vogelsang:1996im}. This difference is crucial to have a consistent calculation
for the polarized structure functions beyond one-loop, as shown in Ref.~\cite{Vogelsang:1996im}. 
From our calculations in the following, we also find that it plays a crucial 
role to fulfill the spin sum rule. After applying dimensional regularization,
we obtain,
\begin{equation}
A^{(1)}_e(0)=\frac{\alpha}{2\pi}\left[-\frac{4}{3}\left(N_\epsilon+\ln\frac{\mu^2}{m^2}\right)-\frac{17}{9}\right]\ .
\end{equation}
From the Ji sum rule, we finally obtain the kinetic electron OAM,
\begin{equation}
L_e=\frac{\alpha}{2\pi}\left[-\frac{2}{3}\left(N_\epsilon+\ln\frac{\mu^2}{m^2}\right)-\frac{10}{9}\right] \ .
\end{equation}
In the following, we will carry out the calculation of OAM in the
Jaffe-Manohar sum rule and compare to the above result.

It is not straightforward to calculate the OAM in the 
Jaffe-Manohar sum rule. We follow the procedure in 
Refs.~\cite{Ji:1996ek,Ji:1995cu} to calculate the matrix element,
\begin{equation}
 \mathcal{L}_e   = {1\over 2P^+}\epsilon_{ij}\lim_{\Delta\to0} {\partial \over \partial i \Delta_{\perp}^i} \langle P'\Lambda|\bar{\psi}(0)
\gamma^+i\partial_\perp^j\psi(0)|P\Lambda\rangle \ ,
\end{equation}
where $\epsilon_{ij}$ is the antisymmetric tensor with $\epsilon_{12}=1$, and 
$\Delta=P'-P$. In the above equation, $\Lambda$ represents the helicity for the electron state.
There is no virtual contribution. In the real contribution, we have to keep the momentum
dependence of $\Delta_\perp$ and perform the derivative in the end to arrive at the
OAM contribution. Again, we will also apply the HVBM scheme for the $\gamma_5$
prescription in the calculations, and the final expression for the OAM can be written 
as,
\begin{eqnarray}
 \mathcal{L}_e &=& \frac{e^2}{2\pi}\int dx\left(1-x^2+\epsilon(1-x)^2\right)\nonumber\\
 &&\times \int \frac{d^{D-2}k_\perp}{(2\pi)^{D-2}}\frac{(1+\epsilon)k_\perp^2}{(k_\perp^2+(1-x)^2m^2)^2} \ ,
\end{eqnarray}
where the $\epsilon$ in the last factor comes from the angular average in $D-2$ dimensions: 
$k_\perp^\alpha k_\perp^\beta\Rightarrow  g_\perp^{\alpha\beta}k_\perp^2/(D-2)$.
In dimensional regularization, we arrive at the following result for ${\cal L}_e$,
\begin{equation}
{\cal L}_e=\frac{\alpha}{2\pi}\left[-\frac{2}{3}\left(N_\epsilon+\ln\frac{\mu^2}{m^2}\right)-\frac{10}{9}\right] \ .
\end{equation}
Indeed, we find that, at the one-loop order, the OAM defined in Jaffe-Manohar and Ji sum rules
agrees with each other, 
\begin{equation}
L_e={\cal L}_e\ .
\end{equation}
Now, let us examine the so-called potential angular momentum contribution. The calculation
follows that for the OAM above, with the replacement of $\partial_\perp^j$ with $A_\perp^j$
in the matrix element. In particular, we will end up with the following integral,
\begin{eqnarray}
&&\int dx (1-x) \int \frac{d^{D-2}k_\perp}{(2\pi)^{D-2}}\nonumber\\
&&\frac{\left(k_\perp-(1-x)\Delta_\perp/2\right)^j}{\left(k_\perp-(1-x)\Delta_\perp/2\right)^2+(1-x)^2m^2
+x(1-x)\Delta_\perp^2/4} \ ,\nonumber\\
\end{eqnarray}
which vanishes in the $k_\perp$ integral. Therefore, at one-loop order, the potential 
OAM does not contribute,
\begin{equation}
l_{\rm pot}=0 \ .
\end{equation}
This proves that the electron OAM defined in Jaffe-Manohar and Ji sum rules 
agrees with each other at one-loop order, and they are consistent with the fact that
the potential OAM vanishes at this order as well.

\subsection{Pauli-Villars regularization}

In this subsection, we check the spin sum rules in the Pauli-Villars regularization. The advantage of this
regularization is that we do not need to worry about the prescription of $\gamma_5$, because the calculations
are performed in 4 dimensions.
As discussed in Ref.~\cite{Burkardt:1990xf}, Pauli-Villars regularization can be consistently implemented in the light-cone gauge by making the following replacement in the photon propagator
\begin{equation}
{id^{\mu\nu}(k)\over k^2} \to {id^{\mu\nu}(k)\over k^2} - {id^{\mu\nu}(k)\over k^2-\Lambda^2} \ ,
\end{equation}
where $d^{\mu\nu}(k)$ takes the following form,
\begin{equation}
d^{\mu\nu}(k)=-g^{\mu\nu}+\frac{n^\mu k^\nu+k^\mu n^\nu}{n\cdot k}\ .
\end{equation}

With this setup, we carry out the calculations with Pauli-Villars regularization for
all the terms discussed in the last subsection, and obtain the following results,
\begin{eqnarray}
{\Delta\Sigma_e\over2} &=& {1\over2} - {\alpha\over 2\pi}\cdot{1\over2}\ ,\\
\mathcal{L}_e &=& - {\alpha\over 2\pi} \left[ {2\over3}\ln{\Lambda^2\over m^2} - {1\over9}\right]\ ,\\
A_e(0)&=& 1 -{\alpha\over 2\pi} \left[ {4\over3}\ln{\Lambda^2\over m^2} + {13\over9}\right]\ ,\\
B_e(0)&=&{\alpha\over 2\pi}\cdot {2\over3} \ ,\\
l_\text{pot}&=& 0 \ .
\end{eqnarray}
The kinetic electron OAM is obtained from the Ji sum rule
\begin{equation}
L_e = {1\over2} [A_e(0)+B_e(0)] - {1\over2}\Delta \Sigma_e = - {\alpha\over 2\pi} \left[ {2\over3}\ln{\Lambda^2\over m^2} - {1\over9}\right] \ .
\end{equation}
Again, we find that the OAM in Jaffe-Manohar and Ji sum rules agrees with each other.
Comparing the above results to those obtained in Ref.~\cite{Burkardt:2008ua}, we find that we agree on $\Delta\Sigma_e$, ${\cal L}_e$, and $B_e(0)$, but
not on $A_e(0)$. In the following section, we will review the light-front wave function
calculation and show that after including a missing contribution, we obtain the
same results as the above. 




\section{Electron Orbital angular momentum from Light-front wave functions}\label{oamlfwf}

In light-front quantization, the physical electron state is expanded in the complete basis of electron and photon Fock states~\cite{Brodsky:1997de}. Up to $O(\alpha)$, the wave function of an electron with momentum $P=(P^+, P^-, \vec{P}_\perp)$ and helicity $\Lambda$ can be expressed as
\begin{eqnarray}
&&|P^+,\vec{P}_\perp, \Lambda\rangle\nonumber\\
&=& \sqrt{Z} \sqrt{2P^+}\, b^\dagger_\Lambda(P,\vec{P}_\perp) |0\rangle \nonumber\\
&& + \sum_{\sigma,h} \int {dxd^2\vec{k}_\perp\over 2(2\pi)^3} 2P^+ \psi^{\Lambda}_{\sigma,h}(x,\vec{k}_\perp)b^\dagger_\sigma(xP^+, x\vec{P}_\perp + \vec{k}_\perp)\nonumber\\
&&\ a^\dagger_h((1-x)P^+,(1-x)\vec{P}_\perp - \vec{k}_\perp) |0\rangle \ ,\nonumber\\
\label{lfwf}
\end{eqnarray}
where $b^\dagger_\sigma$ and $a^\dagger_h$ are the electron and photon creation operators, and $\psi^\Lambda_{\sigma,h}$ is the light-front wave function of the two-particle Fock state. $x$ and $\vec{k}_\perp$ are the light-front momentum fraction and intrinsic transverse momentum of the electron, and the renormalization constant $Z$ ensures that the electron wave function is normalized according to
\begin{eqnarray}
&&\langle P'^+,\vec{P}'_\perp,\Lambda' | P^+, \vec{P}'_\perp, \Lambda\rangle \nonumber\\
&=& 2P^+(2\pi)^3 \delta(P'^+-P^+)\delta^{2}(\vec{P}'_\perp - \vec{P}_\perp)\delta_{\Lambda'\Lambda}\ .
\end{eqnarray}

For the two-particle Fock state, there are four helicity combinations, and their light-front wave functions are~\cite{Brodsky:2000ii}:
\begin{eqnarray}
\psi^\uparrow_{\uparrow,+} &=& {k^1 - ik^2 \over x(1-x)} \phi(x,\vec{k}_\perp) \ ,\nonumber\\
\psi^\uparrow_{\uparrow,-} &=& -{k^1 + ik^2 \over 1-x} \phi(x,\vec{k}_\perp) \ ,\nonumber\\
\psi^\uparrow_{\downarrow,+} &=& {1-x \over x} m \phi(x,\vec{k}_\perp) \ ,\nonumber\\
\psi^\uparrow_{\downarrow,-} &=& 0 \ ,
\end{eqnarray}
and
\begin{eqnarray}
\psi^\downarrow_{\uparrow,+} &=& 0 \ ,\nonumber\\
\psi^\downarrow_{\uparrow,-} &=&{1-x \over x} m \phi(x,\vec{k}_\perp)   \ ,\nonumber\\
\psi^\downarrow_{\downarrow,+} &=& {k^1 - ik^2 \over x(1-x)} \phi(x,\vec{k}_\perp)\ ,\nonumber\\
\psi^\downarrow_{\downarrow,-} &=&  -{k^1 + ik^2 \over 1-x} \phi(x,\vec{k}_\perp)\ ,
\end{eqnarray}
where
\begin{equation}
\phi(x,\vec{k}_\perp) = - {\sqrt{2}e\over \sqrt{1-x}}\, {x(1-x)\over \vec{k}_\perp^2 + (1-x)^2 m^2 + x\lambda^2}\ ,
\end{equation}
with $\lambda$ being the photon mass introduced to regularize potential IR divergences. The coefficients of $\phi(x,\vec{k}_\perp)$ in the above light-front wave functions arise from the following matrix elements
\begin{equation}
{\bar{u}(xP^+, x\vec{P}_\perp + \vec{k}_\perp,\sigma) \over \sqrt{xP^+} }\,  \gamma_\mu \epsilon^{*\mu}_h\, {u(P,\Lambda)\over \sqrt{P^+}} \ ,
\label{spinor}
\end{equation}
where the Dirac spinors are given in Refs.~\cite{Brodsky:2000ii,Lepage:1980fj}.

In light-front quantization, the free-field operators can be expanded as 
\begin{eqnarray}
\psi(\xi) &=& \sum_\sigma \int {dl^+ d^2\vec{l}_\perp \over \sqrt{2l^+}(2\pi)^3)}\nonumber\\
&&\times \left[ b_\sigma u(l,\sigma) e^{-il\cdot \xi} + d^\dagger_\sigma(l) v(l,\sigma) e^{il\cdot\xi} \right]\ ,\nonumber\\
A^\mu(\xi) &=&  \sum_h \int {dl^+ d^2\vec{l}_\perp \over \sqrt{2l^+}(2\pi)^3)}\nonumber\\
&&\times \left[ a_h(l) \epsilon^\mu_h(l) e^{-il\cdot \xi} + a^\dagger_h(l) \epsilon^{*\mu}_h(l) e^{il\cdot\xi} \right]\ ,
\end{eqnarray}
and the free one-particle state $|l\rangle$ is defined as $\sqrt{2l^+} a^\dagger(l) |0\rangle$. Conventions for the photon polarization vector $\epsilon^\mu_h$ have also been given in~\cite{Brodsky:2000ii,Lepage:1980fj}.

With the light-front wave functions above, one can do a direct computation and finds the following wave function renormalization constant
\begin{equation}
Z = 1- \int {dxd^2\vec{k}_\perp \over 2(2\pi)^3}\left(| \psi^\uparrow_{\uparrow,+} |^2 + | \psi^\uparrow_{\uparrow,-} |^2 + | \psi^\uparrow_{\downarrow,+} |^2 \right) \ .
\end{equation}
For an electron with helicity $+1/2$, the electron spin and OAM in the Jaffe-Manohar sum rule can be computed as
\begin{eqnarray}
\Delta\Sigma_e &=& 1- 2 \int {dxd^2\vec{k}_\perp \over 2(2\pi)^3} | \psi^\uparrow_{\downarrow,+} |^2\ ,\nonumber\\
\mathcal{L}_e &=& \int {dxd^2\vec{k}_\perp \over 2(2\pi)^3}(1-x) \left(-| \psi^\uparrow_{\uparrow,+} |^2 + | \psi^\uparrow_{\uparrow,-} |^2\right) \ .\nonumber\\
\label{jaffe}
\end{eqnarray}
On the other hand, 
$A_e(0)$ and $B_e(0)$ can be computed as
\begin{eqnarray}
A_e(0) &=& 1- \int {dxd^2\vec{k}_\perp \over 2(2\pi)^3} (1-x)\nonumber\\
&&\times\left(| \psi^\uparrow_{\uparrow,+} |^2 + | \psi^\uparrow_{\uparrow,-} |^2 + | \psi^\uparrow_{\downarrow,+} |^2 \right) \ ,\nonumber\\
B_e(0) &=& \int {dxd^2\vec{k}_\perp \over 2(2\pi)^3} (1-x)m^2|\phi(x,\vec{k}_\perp)|^2\ ,
\label{ji}
\end{eqnarray}
which are the same as the results of Burkardt et al.~\cite{Burkardt:2008ua}. Clearly, there are UV divergences in the above equations, for which 
we will apply the Pauli-Villars regularization. Before we proceed to that, we would like to comment on dimensional regularization 
calculations of the above integrals in Ref.~\cite{Liu:2014fxa}. Because of the divergences, we have to keep the 
$\epsilon$-terms in the light-front wave functions if we want to apply dimensional regularization 
in the above equations. The naive implementation as that of Ref.~\cite{Liu:2014fxa} will miss finite contributions. 
Compared to the explicit expressions from the above equations with $x$ and $k_\perp$ dependence to the results in the previous section, 
Eqs.~(12,16,20), we will find that they agree except for all the $\epsilon$-terms. Without the $\epsilon$-terms,
dimensional regularization will not give the complete results. 

In the Pauli-Villars regularization, instead, we perform the calculation in $4$ dimensions. For each divergent $k_\perp$ integral, there is a Pauli-Villars subtraction with $\lambda^2\to\Lambda^2$, and the $\lambda^2\to0, \Lambda^2\gg m^2$ limit is taken at the end of the calculation. In this way, the results for Eqs.~(\ref{jaffe}--\ref{ji}) are
\begin{eqnarray}
\Delta\Sigma_e &=& 1 - {\alpha\over 4\pi}\cdot 2\ ,\label{pvspin}\\
\mathcal{L}_e &=& - {\alpha\over 4\pi} \left[ {4\over3}\ln{\Lambda^2\over m^2} - {2\over9}\right]\ ,
\label{pvoam}
\end{eqnarray}
and
\begin{eqnarray}
A_e(0)&=& 1 -{\alpha\over 4\pi} \left[ {8\over3}\ln{\Lambda^2\over m^2} + {44\over9}\right]\ ,\label{pva}\\
B_e(0)&=&{\alpha\over 4\pi}\cdot {4\over3} \ .
\label{pvb}
\end{eqnarray}
The kinetic electron OAM is obtained from the Ji sum rule
\begin{equation}
L_e = {1\over2} [A_e(0)+B_e(0)] - {1\over2}\Delta \Sigma_e = - {\alpha\over 4\pi} \left[ {4\over3}\ln{\Lambda^2\over m^2} + {7\over9}\right] \ .
\end{equation}
From this result, it was then concluded in Ref.~\cite{Burkardt:2008ua} that
\begin{equation}
\mathcal{L}_e - L_e = {\alpha\over4\pi} \neq 0 \ ,
\label{difference}
\end{equation}
where the authors also attributed this discrepancy between the canonical and kinetic OAM to the potential angular momentum. Later on, Liu and Ma~\cite{Liu:2014fxa} calculated the potential angular momentum to $\mathcal O(\alpha)$, and found that it has a non-vanishing anomalous dimension. As already mentioned in the Introduction, this cannot be correct because the potential angular momentum does not contribute at this order to the evolution of angular momentum~\cite{Ji:1996ek, Ji:1995cu}. Actually, our direct evaluation of the matrix element of $l_\text{pot}$ yields zero (see Appendix~\ref{lcwfpot}).

Liu and Ma~\cite{Liu:2014fxa} also considered the surface terms in the electron angular momentum density operator for the Ji sum rule
\begin{eqnarray}
M^{\mu\nu\rho}_\text{surf} &=& - {i\over2} \left(x^\nu \partial^\rho - x^\rho \partial^\nu\right) (\bar{\psi} \gamma^\mu \psi) \nonumber\\
&&+ {1\over4} \partial_\rho \left[\left(x^\nu\epsilon^{\mu\lambda\rho\sigma}- x^\lambda\epsilon^{\mu\nu\rho\sigma}\right)\bar{\psi}\gamma_\sigma\gamma^5\psi\right] \ .
\end{eqnarray}
While the contribution from these surface terms is claimed by them to be non-zero, our direct evaluation shows that their off-forward matrix elements vanish in the forward limit.


With zero contributions from the potential angular momentum and surface terms, there seems to be an inconsistency between the Jaffe-Manohar and Ji sum rules from Eq.~(\ref{difference}), at least at $\mathcal O(\alpha)$ in QED. However, as we now explain, this is because the calculation by Burkardt et al.~\cite{Burkardt:2008ua} is not complete. In order to see what is missing in their calculation, it is worthwhile to recall the difference between light-front quantization and equal-time quantization in the light-cone gauge. In the equal-time quantization, the light-cone gauge photon propagator is given by $id^{\mu\nu}(k)/k^2$ with
\begin{eqnarray}
d^{\mu\nu}(k) &=& - g^{\mu\nu} + {n^\mu k^\nu + n^\nu k^\mu \over n\cdot k} \nonumber\\
&=& \sum_{h=\pm} \epsilon^{*\mu}_h(k) \epsilon^\nu_h(k) + {k^2n^\mu n^\nu \over (n\cdot k)^2} \ ,
\label{equaltime}
\end{eqnarray}
which means that both the transverse and longitudinal modes propagate. In contrast, in light-front quantization at equal light-front time, one has~\cite{Srivastava:2000cf}
\begin{eqnarray}
\tilde{d}^{\mu\nu}(k) &=& - g^{\mu\nu} + {n^\mu k^\nu + n^\nu k^\mu \over n\cdot k} - {k^2 n^\mu n^\nu \over (n\cdot k)^2}\nonumber\\
&=& \sum_{h=\pm} \epsilon^{*\mu}_h(k) \epsilon^\nu_h(k)\ ,
\label{equallc}
\end{eqnarray}
Since the photon is essentially put on-shell in the latter case, it is not surprising that only the physical transverse modes propagate. If the photon is massless, the difference between the above two equations vanishes. In general, the extra term proportional to $k^2$ in Eq.~(\ref{equallc}) is compensated for by an instantaneous interaction in the light-front Hamiltonian, which comes from eliminating the non-dynamical field $A^-$ in terms of the dynamical ones in the light-cone gauge (there is also an instantaneous fermion propagator in light-front quantization, but it does not contribute here)~\cite{Srivastava:2000cf}. In this way, physics remains the same in light-front and equal-time perturbation theories.

From the discussions above, we can see that for a massive photon, the contribution of the instantaneous interaction does not vanish in the light-front wave functions arising from the matrix elements in Eq.~(\ref{spinor}), and therefore has to be taken into account. Its contribution is equivalent to adding an extra independent polarization vector,
\begin{equation}
\epsilon^{\mu}_0(k) = - {\sqrt{k^2}n^\mu\over n\cdot k} \ ,
\end{equation}
which results in extra light-front wave functions for the two-particle state,
\begin{equation}
\psi^\uparrow_{\uparrow,0}(x,\vec{k}_\perp) = \psi^\downarrow_{\downarrow,0}(x,\vec{k}_\perp) = - {\sqrt 2\lambda \over (1-x) } \phi(x,\vec{k}_\perp) \ .
\label{extralcwf}
\end{equation}
If one adopts the Pauli-Villars regularization, they will contribute to $\Delta\Sigma_e$ and $A_e(0)$ of the Pauli-Villars particle with $\lambda$ replaced by $\Lambda$ in Eq.~(\ref{extralcwf}). In the $\lambda\to0, \Lambda\gg m$ limit, their contributions to the real and virtual parts of $\Delta\Sigma_e$ cancel, while $A_e(0)$ receives a nonzero correction,
\begin{equation}
\delta A_e(0) = \int {dxd^2\vec{k}_\perp \over 2(2\pi)^3} (1-x) |\psi^\uparrow_{\uparrow,0}|^2 = {\alpha\over 2\pi}\ , 
\label{deltaL}
\end{equation}
which exactly cancels the difference between the canonical and kinetic OAM in Eq.~(\ref{difference}).


In summary, there is no difference between the canonical and kinetic OAM for an electron at $\mathcal O(\alpha)$ in QED. The perturbative calculation with light-front wave functions is subtle due to the existence of an extra instantaneous interaction in the light-front Hamiltonian. It is the contribution of this term that is missing and leads to the incomplete result of Burkardt et al.~\cite{Burkardt:2008ua} and Liu and Ma~\cite{Liu:2014fxa} in the Pauli-Villars regularization. 

\section{Conclusion}\label{concl}

In this paper, we reexamined the canonical and kinetic OAM for a dressed electron at $\mathcal O(\alpha)$ in QED. We performed the calculation both with Feynman diagrams and with light-front wave functions, and in the former case we used both dimensional regularization and Pauli-Villars regularization for the UV divergences. Our results show that the canonical and kinetic electron OAM are equal, and the contribution from the potential angular momentum is zero. The reason that Burkardt et al.~\cite{Burkardt:2008ua} and Liu and Ma~\cite{Liu:2014fxa} obtained a finite difference between the canonical and kinetic electron OAM is that the contribution of an instantaneous interaction term is missing in their computation using light-front wave functions and Pauli-Villars regularization. After including its contribution, the Jaffe-Manohar and Ji sum rule are then consistent with each other.


It should also be pointed out that our result is at $\mathcal O(\alpha)$ in QED. When going to higher perturbative orders, it is likely that the potential angular momentum will start to contribute, but the Jaffe-Manohar and Ji sum rules are expected to be consistent with each other.

\vspace{2em}
This work was partially supported by the U.S. Department of Energy Office of Science, Office of Nuclear Physics under Award Number DE-FG02-93ER-40762 and DE-AC02-05CH11231,
and a grant (No. 11DZ2260700) from the Office of
Science and Technology in Shanghai Municipal Government, and also by grants from the National Science Foundation of China (No. 11175114, No. 11405104) and by a DFG grant SCHA 458/20-1.

\vspace{1cm}

\appendix

\section{Light-front wave function calculation of the potential angular momentum}
\label{lcwfpot}

The potential angular momentum can be expressed as
\begin{equation}
\langle l_{\text{pot}}^z\rangle =-{e\over 2P^+}\epsilon_{ij}  \lim_{P'_\perp\to0} {\partial\over i\partial P'^i_\perp} \langle P' \Lambda | \bar{\psi}(0) \gamma^+ A^j(0) \psi(0) | P\Lambda\rangle \ ,
\label{potam}
\end{equation}
where we have chosen $\vec{P}_\perp =0$.

From the light-front wave functions, we have
\begin{eqnarray}
 &&\langle P',\uparrow| \bar{\psi}(0) \gamma^+ A^j(0) \psi(0) | P,\uparrow\rangle \nonumber\\
 &=&  e \sqrt{Z}\sum_{\sigma,h}\int {dxd^2\vec{k}_\perp\over 2(2\pi)^3} {1\over \sqrt{x(1-x)}} \psi^+_{\sigma,h}(x,\vec{k}_\perp)\nonumber\\
 &&\times \bar{u}(P',\uparrow) \gamma^+ u_\sigma(xP^+, x\vec{P}_\perp + \vec{k}_\perp)\nonumber\\
 &&\times \epsilon^j_h((1-x)P^+, (1-x)\vec{P}_\perp - \vec{k}_\perp)\nonumber\\
 &&+   e \sqrt{Z}\sum_{\sigma,h}\int {dxd^2\vec{k}_\perp\over 2(2\pi)^3} {1\over \sqrt{x(1-x)}} \psi^{+,*}_{\sigma,h}(x,\vec{k}_\perp)\nonumber\\
 &&\times \bar{u}_\sigma(xP'^+, x\vec{P}'_\perp + \vec{k}_\perp) \gamma^+ u(P,\uparrow)\nonumber\\
 &&\times \epsilon^{*j}_h((1-x)P'^+, (1-x)\vec{P}'_\perp - \vec{k}_\perp)\ .
 \label{lcpot2}
\end{eqnarray}
Using the explicit expressions of the electron spinor and photon polarization vector in Refs.~\cite{Brodsky:2000ii,Lepage:1980fj}, we have
\begin{equation}
\bar{u}(p',\sigma') \gamma^+ u(p,\sigma) = 2\sqrt{p'^+p^+}\delta_{\sigma\sigma'}\ ,
\end{equation}
and
\begin{equation}
\vec{\epsilon}_\perp^\pm = \mp {\hat{x} \pm i\hat{y} \over \sqrt{2}}\ .
\end{equation}
Plugging them into Eq.~(\ref{lcpot2}), 
we find that the matrix element of interest has no dependence on $\vec{P}'_\perp$, 
so that the potential angular momentum in Eq.~(\ref{potam}) must be zero.

\end{document}